\documentclass[runningheads]{llncs}
\usepackage{multirow}
\usepackage{xpatch}  
\usepackage{tabu}  
\usepackage{makecell}
\usepackage[misc]{ifsym}

\usepackage{graphicx}
\usepackage{hyperref}

\begin{document}
\title{BiOnt: Deep Learning using Multiple Biomedical Ontologies for Relation Extraction\thanks{This work was supported by FCT through project  DeST: Deep Semantic Tagger, ref.\ PTDC/CCI-BIO/28685/2017, LASIGE Research Unit, ref.\ UIDP/00408/2020, and PhD Scholarship, ref.\ SFRH/BD/145221/2019.}}
\titlerunning{BiOnt}
%
\author{Diana Sousa\textsuperscript{\Letter}\orcidID{0000-0003-0597-9273} \and
Francisco M. Couto\orcidID{0000-0003-0627-1496}}
\authorrunning{D. Sousa and F. M. Couto}
\institute{LASIGE, Faculdade de Ci\^{e}ncias, Universidade de Lisboa, Portugal \email{dfsousa@lasige.di.fc.ul.pt}}
\maketitle              
\begin{abstract}
Successful biomedical relation extraction can provide evidence to researchers and clinicians about possible unknown associations between biomedical entities, advancing the current knowledge we have about those entities and their inherent mechanisms. Most biomedical relation extraction systems do not resort to external sources of knowledge, such as domain-specific ontologies. However, using deep learning methods, along with biomedical ontologies, has been recently shown to effectively advance the biomedical relation extraction field. To perform relation extraction, our deep learning system, BiOnt, employs four types of biomedical ontologies, namely, the Gene Ontology, the Human Phenotype Ontology, the Human Disease Ontology, and the Chemical Entities of Biological Interest, regarding gene-products, phenotypes, diseases, and chemical compounds, respectively. We tested our system with three data sets that represent three different types of relations of biomedical entities. BiOnt achieved, in F-score, an improvement of 4.93 percentage points for drug-drug interactions (DDI corpus), 4.99 percentage points for phenotype-gene relations (PGR corpus), and 2.21 percentage points for chemical-induced disease relations (BC5CDR corpus), relatively to the state-of-the-art. The code supporting this system is available at \   \url{https://github.com/lasigeBioTM/BiONT}.

\keywords{Relation Extraction \and Biomedical Ontologies \and Deep Learning \and Text Mining}
\end{abstract}
\section{Introduction}

The description of the mechanisms that are responsible for the behavior of biological systems is non-trivial, and each step towards the understanding of those mechanisms often constitutes a scientific achievement \cite{YU2006252,BECHTEL2007269}. Typical examples describe diseases that are associated with mechanisms that originate phenotypic abnormalities as a result of modified gene expression, as well as the action of drugs on those diseases \cite{Campaner2011}, among others. One significant step to fully understand biological systems mechanisms is to extract and classify the relations that exist between the different biomedical entities, namely chemicals, diseases, genes, and phenotypes. In literature, authors classify this problem as a Relation Extraction (RE) task. Biomedical RE aims to extract and classify relations between biomedical entities in highly heterogeneous or unstructured scientific or clinical text.

Deep learning is widely used to solve problems such as speech recognition, visual object recognition, and object detection. Lately, deep learning based-systems have started to tackle RE problems. These systems are becoming increasingly more complex, namely the MIMLCNN \cite{jiang-etal-2016-relation}, and PCNN + Att \cite{Lin2016NeuralRE} systems, that mark recent turning points in the deep learning RE field. Both of these systems use Word2Vec \cite{NIPS2013_5021} that aims to capture the syntactic and semantic information about the word \cite{journals/corr/Kumar17a}. However, deep learning methods that effectively extract and classify relations between biomedical entities in the text are still scarce \cite{Li2017ANJ,Lamurias2019}. 

Ontologies play an important role in biomedical research through a variety of applications and are used primarily as a source of vocabulary for standardization and integration purposes \cite{Bodenreider08biomedicalontologies}. Word embeddings can learn how to detect relations between entities but manifest difficulties in grasping the semantics of each entity and their specific domain. Domain-specific ontologies provide and formalize this knowledge. Thus, a structured representation of the semantics between entities and their relations, an ontology, allows us to use it as an added feature to a machine learning classifier. Some of the biomedical entities structured in publicly available ontologies are genes properties/attributes (Gene Ontology (\textbf{GO})) (45003 terms) \cite{GO,10.1093/nar/gky1055}, phenotypes (Human Phenotype Ontology (\textbf{HPO})) (25810 terms) \cite{HPO}, diseases (Human Disease Ontology (\textbf{DO})) (18114 terms) \cite{10.1093/nar/gky1032}, and drugs/chemicals (Chemical Entities of Biological Interest (\textbf{ChEBI})) (133104 terms) \cite{CHEBI}\footnote{term counts at \textit{09/09/2019}}.  

This work presents the \textbf{BiOnt} system, a biomedical RE system built using bidirectional Long Short-Term Memory (LSTM) networks. The BiOnt system incorporates the state-of-the-art Word2Vec word embeddings \cite{NIPS2013_5021} and makes use of different combinations of input channels to maximize performance. Our system is based on the work of Lamurias et al. \cite{Lamurias2019} and Xu et. al. \cite{8377981}. Both of these models make use of biomedical resources as embedding layers for their respective systems. Lamurias et. al. \cite{Lamurias2019} uses the Xu et. al. \cite{8377981} model has a baseline with an added ontological embedding layer (BO-LSTM model). However, the BO-LSTM model is limited to two types of relations, namely, drug-drug, and gene-phenotype relations.

\begin{figure}[ht]
\centering
\includegraphics[width=9cm]{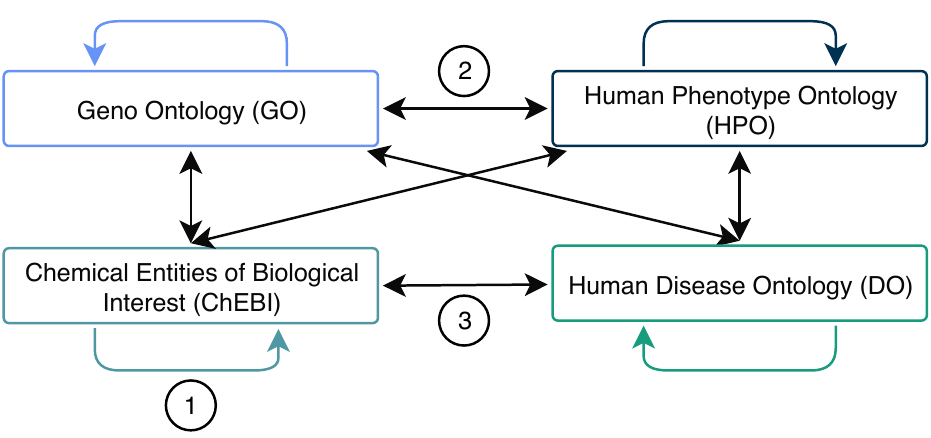}
\caption{The ten possible combinations between the four biomedical ontologies (the Gene Ontology (GO) \cite{GO}, the Human Phenotype Ontology (HPO) \cite{HPO}, the Human Disease Ontology (DO) \cite{10.1093/nar/gky1032}, and the Chemical Entities of Biological Interest (ChEBI) \cite{CHEBI}). The \textbf{1} represents the DDI corpus, the \textbf{2} the PGR corpus, and the \textbf{3} the BC5CDR corpus (described in Section 3).} \label{fig1}
\end{figure}

External sources of knowledge, such as biomedical ontologies, can provide highly valuable information for the detection of relations between entities in the text, as described previously by Lamurias et. al. \cite{Lamurias2019}. These knowledge-bases provide not only relevant characteristics about the respective entities but also about the underlying semantics of the relations they establish. This information is not expressed directly in the training data but usually reinforces a relation between two entities in the text. The novelty of our system is that expands the previous work done by Lamurias et. al. \cite{Lamurias2019} by using four types of domain-specific ontologies, and combine them to extract new types of relations, along with word embeddings \cite{NIPS2013_5021} and WordNet hypernyms \cite{Ciaramita:2006:BSD:1610075.1610158}. BiOnt successfully replicates the results of the BO-LSTM application, using different types of ontologies. Our system can extract new relations between four biomedical entities, namely, genes, phenotypes, diseases, and chemicals. Figure \ref{fig1} shows how these four types of biomedical ontologies can be combined to aid the relation extraction of ten different combinations of biomedical entities. The BiOnt system also explores the use of entities that are not direct entries in an ontology (e.g., genes), linking each entity to their most informative annotation concept within a corresponding ontology (e.g., GO). Our method incorporates more ontologies than the previously mentioned systems and is evaluated using three state-of-the-art data sets. The BiOnt system can be used to effectively populate knowledge bases regarding gold standard relations. Ultimately, it can be used to explore new experimental hypotheses providing evidence to researchers and clinicians about possible unknown associations between biomedical entities.

\vspace{-1cm}

\section{Methodology}

This section describes the BiOnt model with an emphasis on the enhancements done to BO-LSTM \cite{Lamurias2019} model to allow multi-ontology integration, expanding the number of different type candidate pairs from two to ten. The BiOnt model uses a combination of different language and entity related data representations, that feed individual channels creating a multichannel architecture. The input data is used to generate instances to be classified by the model. Each instance corresponds to a candidate pair of entities in a sentence. To each instance, the model assigns a positive or negative class. A positive class corresponds to an identified relation between two biomedical concepts, where the nature of this relation depends of the data set being used to perform the evaluation, and a negative class implies no relation between the different entities. 

\begin{figure}
\centering
\includegraphics[width=11cm]{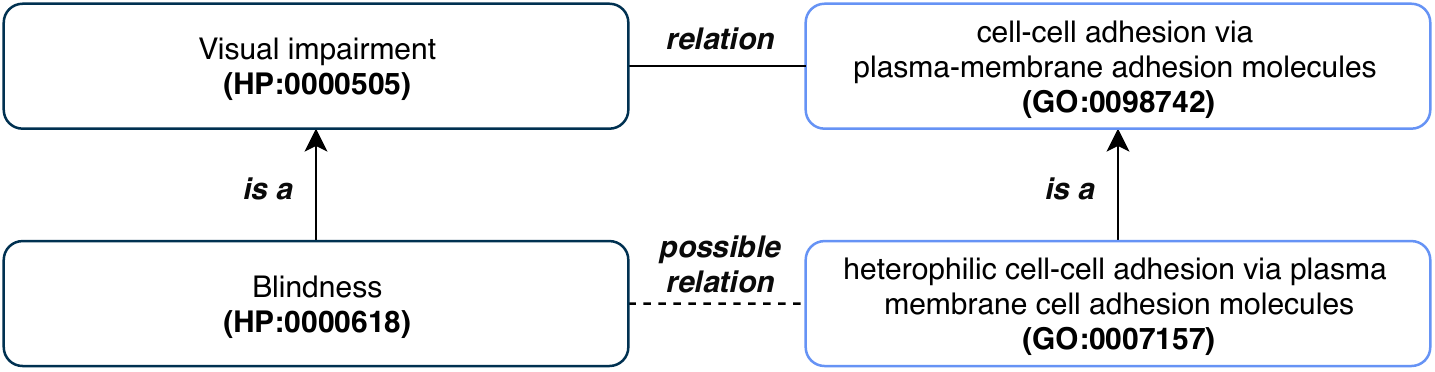}
\caption{BiOnt ontology embedding illustration based on the HPO and the GO ontologies, for the candidate relation between the human phenotype \textit{blindness} and the gene \textit{CRB1} (represented by the GO term \textit{heterophilic cell-cell adhesion via plasma membrane cell adhesion molecules}).} \label{fig6}
\end{figure}

An instance should condense all relevant information to classify a candidate pair. To create an instance the BiOnt model relies on three primary data information layers. After sentence tokenization, these layers are: Shortest Dependency Path (SDP) \cite{Pyysalo:2013b,Mikolov:2013:DRW:2999792.2999959}, WordNet Classes \cite{Ciaramita:2006:BSD:1610075.1610158}, and \textbf{Ontology Embeddings}. The latter represents the relations between the ancestors for each ontology concept corresponding to an entity (Figure \ref{fig6}). The model assumes that the input data already has the offsets of the relevant entities identified and their respective concept ID, the Named-Entity Recognition and Linking tasks. However, while most entities already corresponded to an ontology concept ID, some entities, such as genes do not have a direct entry in an ontology. BiOnt matches these entities to their most representative concept in the Gene Ontology \cite{GO}. To match the gene to their most representative GO term the priority was given to concepts inferred from experiments, for having a more sustained background and usually be more descriptive. For tie-breaking, if we have several GO terms inferred from experiments, the choice is the term that is the most specific (i.e., with the longer ancestry line).

\vspace{-0.3cm}

\vspace{-0.5cm}

As stated previously, our system expands the work done by Lamurias et. al. \cite{Lamurias2019} by using four types of domain-specific ontologies, and combine them to extract new types of relations. Therefore, to allow this diversity of relations, we adapted the BO-LSTM model common ancestors and the concatenation of ancestors channels. Since the common ancestors' channel could only be used for relations between the same type of biomedical entities, we only use the concatenation of ancestors channel for the relations between different biomedical entities. 

\vspace{-0.3cm}

\section{Evaluation}

To showcase our systems' performance, we used three different state-of-the-art data sets. These data sets represent three out of the ten possible combinations of the biomedical entities used in this work, drug-drug interactions, phenotype-gene relations, and chemical-induced disease relations. With these data sets, we intend to show the flexibility of our model to the different types of biomedical entities represented by biomedical ontologies. Figure \ref{fig1} illustrates how the entities present in the three data sets (\textbf{1}, \textbf{2}, and \textbf{3}) are connected to the different biomedical entities.

\textbf{Drug-Drug Interactions (1)} The SemEval 2013: Task 9 DDI Extraction Corpus \cite{HERREROZAZO2013914} is a corpus that describes drug-drug interactions (DDIs) focused on both pharmacokinetic (PK) and pharmacodynamic (PD) DDIs. The manually annotated corpus created by Herrero et. al. \cite{HERREROZAZO2013914} combines 5028 DDIs, from selected texts of the DrugBank database and  Medline abstracts. 

\textbf{Phenotype-Gene Relations (2)} The Phenotype-Gene Relations Corpus (PGR) \cite{sousa-etal-2019-silver} is a corpus that describes human phenotype-gene relations, created in a fully automated manner. Due to being a silver standard corpus is not expected to be as reliable as manually annotated corpora. Nonetheless, the authors show the system efficiency by training two state-of-the-art relation extraction deep learning systems. The PGR corpus combines 4283 human phenotype-gene relations. 

\textbf{Chemical-Induced Disease Relations (3)} The BioCreative V CDR Corpus (BC5CDR) \cite{10.1093/database/baw068} is a corpus of chemical-induced disease (CID) relations. The BC5CDR corpus consists of 3116 chemical-disease interactions annotated from PubMed articles. To use the BC5CDR corpus, we had to preprocess the documents linking the annotations of the relations to their sentences. We assumed that if two entities share a relation in the document, they will continue to share that relation if present in the same sentence of that document.


\section{Results and Discussion}

Table \ref{tab2} presents the relation extraction results of our system, BiOnt, for each data set. For all three data sets, our system performs better using the ontological embeddings layer (+ Ontologies), than just using the word embeddings and WordNet classes layers (State-of-the-art), by an average of 0.0404. The most relevant contribution for this metric was an increase in recall for the DDI and PGR corpus, and in precision for the BC5CDR corpus. The ontology embeddings contribute to the identification of more correct relations, with a small trade-off in precision, for the DDI corpus. For the other two data sets, the ontological embedding layer does not damage the precision, while more correct relations are identified.

\begin{table}
\renewcommand\arraystretch{1.3}
\centering
\caption{Relation extraction results with the BiOnt system, for each data set, expressing drug-drug interactions (DDI Corpus), phenotype-gene relations (PGR Corpus), and chemical-induced disease relations (BC5CDR Corpus).}\label{tab2}
\newcolumntype{C}{ >{\centering\arraybackslash} m{3cm} }
\newcolumntype{A}{ >{\centering\arraybackslash} m{1.7cm} }
\newcolumntype{B}{ >{\centering\arraybackslash} m{2cm} }
\begin{tabular}{|C|C|A|A|A|}
\hline
\textbf{Data Set} & \textbf{Configuration} & \textbf{Precision} & \textbf{Recall} & \textbf{F-score} \\
\Xhline{2\arrayrulewidth}
\multirow{2}{*}{DDI Corpus} & State-of-the-art & 0.7134 & 0.6410 & 0.6753 \\
\cline{2-5}
&  + Ontologies & 0.6784 & 0.7775 & \textbf{0.7246} \\
\hline\hline
\multirow{2}{*}{PGR Corpus} & State-of-the-art & 0.8421 & 0.6666 & 0.7442 \\
\cline{2-5}
& + Ontologies & 0.8438 & 0.7500 & \textbf{0.7941} \\
\hline\hline
\multirow{2}{*}{BC5CDR Corpus} & State-of-the-art & 0.5371 & 0.7264 & 0.6175 \\
\cline{2-5}
& + Ontologies & 0.5770 & 0.7173 & \textbf{0.6396} \\
\hline
\end{tabular}
\end{table}

For the DDI corpus, the BiOnt system, due to the inherent variability of the preprocessing phase (by randomizing the division between training and test sets), when comparing with the BO-LSTM system, performed slightly worse (0.7246 in F-score) than the previously reported results (0.7290 in F-score).
The paper supporting the PGR corpus \cite{sousa-etal-2019-silver} reported some deep learning applications results, including with the BERT \cite{devlin-etal-2019-bert} based BioBERT \cite{BIOBERT} pre-trained biomedical language representation model (0.6716 in F-score). Our system outperformed those results with an F-score of 0.7941.
Regarding the BC5CDR corpus, our system outperformed the best system (0.5703 in F-score) in the challenge task chemical-induced disease (CID) relation extraction of BioCreative V, by 0.0693 \cite{inproceedingscdr}, with 0.6396 in F-score. 
The differences in F-score, for the distinct data sets, are mostly due to how they were built, and the completeness and complexity of the respective ontologies. 
For instance, the PGR corpus is a silver standard corpus, therefore, could have entities that were poorly identified, not identified at all, or not linked to the right identifier. The BC5CDR corpus was annotated for documents, not regarding the offsets of the entities that shared a relation in each document, which is also a possible limitation.

\vspace{-0.2cm}

\section{Conclusions and Future Work}

This work showed that the knowledge encoded in biomedical ontologies plays a vital part in the development of learning systems, providing semantic and ancestry information for entities, such as genes, phenotypes, chemicals, and diseases. We evaluated BiOnt using three state-of-the-art data sets (DDI, PGR, and BC5CDR corpus), obtaining improvements in F-score (4.93, 4.99, and 2.21 percentage points, respectively), by using an ontological information layer. Our system successfully enhances the results of Lamurias et. al. \cite{Lamurias2019} to other entities and ontologies. BiOnt shows that integrating biomedical ontologies instead of relying solely on the training data for creating classification models will allow us not only to find relevant information for a particular problem quicker but possibly also to find unknown associations between biomedical entities.

Regarding future work, it is possible to integrate more ontological information, and in different ways. For instance, one could consider only the relations between the ancestors with the highest information content (more relevant for the candidate pair they characterize). The information content could be inferred from the probability of each term in each ontology or resorting to an external data set. Also, a semantic similarity measurement could account for non-transitive relations (within the same ontology). Relatively to biomedical concepts that do not constitute ontology entries, we could explore quantitative evidence values, choose more than one representative term, and we could also employ semantic similarity measures \cite{COUTO2019870}. 




%
%

%
%
\bibliographystyle{splncs04}
\bibliography{bibliography}
\end{document}